\documentclass[prl,aps,twocolumn,showpacs,superscriptaddress]{revtex4}

\usepackage{natbib}
\usepackage{amssymb}
\usepackage{graphicx}

\begin{document}

\title{Pressure Induced Stripe-order Antiferromagnetism and First-order Phase Transition in FeSe}

\author{P. S. Wang}
\affiliation{Department of Physics and Beijing Key Laboratory of Opto-electronic Functional Materials $\&$ Micro-nano Devices, Renmin University of China, Beijing, 100872, China}
\author{S. S. Sun}
\affiliation{Department of Physics and Beijing Key Laboratory of Opto-electronic Functional Materials $\&$ Micro-nano Devices, Renmin University of China, Beijing, 100872, China}
\author{Y. Cui}
\affiliation{Department of Physics and Beijing Key Laboratory of Opto-electronic Functional Materials $\&$ Micro-nano Devices, Renmin University of China, Beijing, 100872, China}
\author{W. H. Song}
\affiliation{Department of Physics and Beijing Key Laboratory of Opto-electronic Functional Materials $\&$ Micro-nano Devices, Renmin University of China, Beijing, 100872, China}
\author{T. R. Li}
\affiliation{Department of Physics and Beijing Key Laboratory of Opto-electronic Functional Materials $\&$ Micro-nano Devices, Renmin University of China, Beijing, 100872, China}
\author{Rong Yu}
\email{rong.yu@ruc.edu.cn}
\affiliation{Department of Physics and Beijing Key Laboratory of Opto-electronic Functional Materials $\&$ Micro-nano Devices, Renmin University of China, Beijing, 100872, China}
\affiliation{Department of Physics and Astronomy, Shanghai Jiao Tong University, Shanghai 200240, China and Collaborative Innovation Center of Advanced Microstructures, Nanjing %%@
210093, China}
\author{Hechang Lei}
\email{hlei@ruc.edu.cn}
\affiliation{Department of Physics and Beijing Key Laboratory of Opto-electronic Functional Materials $\&$ Micro-nano Devices, Renmin University of China, Beijing, 100872, China}

\author{Weiqiang Yu}
\email{wqyu_phy@ruc.edu.cn}
\affiliation{Department of Physics and Beijing Key Laboratory of Opto-electronic Functional Materials $\&$ Micro-nano Devices, Renmin University of China, Beijing, 100872, China}
\affiliation{Department of Physics and Astronomy, Shanghai Jiao Tong University, Shanghai 200240, China and Collaborative Innovation Center of Advanced Microstructures, Nanjing %%@
210093, China}

\date{\today}


\begin{abstract}
To elucidate the magnetic structure and the origin of the nematicity in FeSe, we perform a high-pressure $^{77}$Se NMR study on FeSe single crystals. We find a suppression of the %%@
structural transition temperature with pressure up to about 2 GPa from the anisotropy of the Knight shift. Above 2 GPa, a stripe-order antiferromagnetism that breaks the spatial %%@
four-fold rotational symmetry is determined by the NMR spectra under different field orientations and with temperatures down to 50 mK. The magnetic phase transition is revealed to be %%@
first-order type, implying the existence of a concomitant structural transition via a spin-lattice coupling. Stripe-type spin fluctuations are observed at high temperatures, and %%@
remain strong with pressure. These results provide clear evidences for strong coupling between nematicity and magnetism in FeSe, and therefore support a universal scenario of magnetic %%@
driven nematicity in iron-based superconductors.
\end{abstract}
\maketitle

In most iron-based superconductors, superconductivity emerges near an antiferromagnetic
(AFM) phase, making it important to study the nature of their magnetism~\cite{NatP_6_645,Stewart_RevMP,P_Dai_RevMP}. For iron pnictides, the
magnetic ground state typically has a stripe-type, or ($\pi$, 0), AFM order~\cite{P_Dai_Nat}. The magnetic transition at $T_N$ is preceded by a
tetragonal-to-orthorhombic structural transition at $T_s$ $\geq$ $T_N$ where the lattice $C_4$ symmetry is broken, and the orthorhombic phase
is denoted as a nematic state. The nematicity can be observed as anisotropy in the in-plane resistivity~\cite{Science_329_824}
and spin fluctuations~\cite{P_Dai_Nat}, and the splitting of the degenerate $d_{xz}$/$d_{yz}$ orbitals~\cite{Yi11} (so called orbital ordering~\cite{Lv}).
Although it is generally believed that the nematicity has an electronic origin, it is still highly debated whether
the nematicity is driven by the spin fluctuations or the orbital ordering~\cite{Rev_nematic}.

Recent discoveries in the bulk FeSe superconductors~\cite{WuMK_PNAS_2008}  make this unsolved issue even more elusive: at ambient pressure, the electronic nematicity shows up below %%@
$T_s$ $\sim$ 90 K~\cite{Cava_PRL_2009,Shimojima_PRB}, while a magnetic ordered state is absent. Applying pressure above 1 GPa, however, magnetic ordering emerges, and the ordering %%@
temperature $T_N$ increases with pressure~\cite{Imai_PRL_2009,Bendele_uSR_PRL_2010}. Meanwhile, $T_s$ is substantially suppressed at $\sim$ 2 GPa~\cite{Terashima_JPSJ_2015,SunJP_NC}. %%@
These seemingly sharp contrasts between FeSe and iron pnictides challenge the existed view of the interplay among nematicity, orbital ordering and magnetism, and inspire various %%@
theoretical proposals for the nature of magnetism in FeSe~\cite{PRL_RongYu,Nat_Phy_WangFa,Nat_Phy_Frustrat,LiuK_PRB_2016}. On the experimental side, inelastic neutron scattering %%@
measurements suggest the coexistence of stripe and checkerboard spin fluctuations at ambient pressure~\cite{Rahn_PRB_2015,ZhaoJun_arxiv_1511}, and transport measurements report %%@
high-temperature superconductivity (HTSC) in both the ambient-pressure nematic and the high-pressure magnetic %%@
phases~\cite{Terashima_JPSJ_2015,SunJP_NC,Garbarino_34K_EPL,Medvedev_NatMat_HP}. Therefore, resolving the nematicity and the magnetic structure in FeSe not only helps building up the %%@
proper theory on the magnetism of FeSe, but also becomes important in understanding the HTSC in FeSe and other iron-based superconductors.

\begin{figure}[t]
\includegraphics[bb=0 0 864 555, width=7.5cm, height=5.3cm]{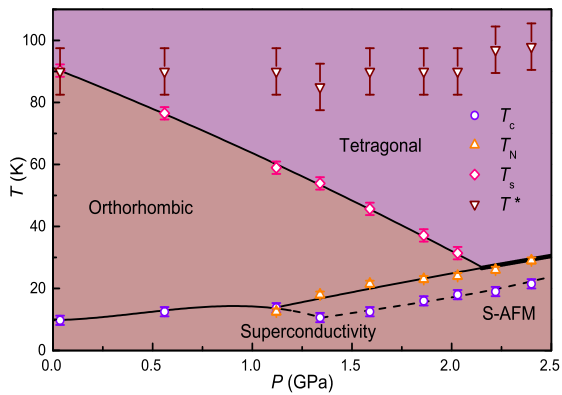}
\caption{\label{Fig1}(color online). The ($P$, $T$) phase diagram with $T_s$, $T_N$, $T_c$, and $T^*$ (see text for definition). The S-AFM phase is observed below $T_N$ above 2 GPa. %%@
The first-order magnetic transition is indicated by the bold solid line of $T_N$. The solid line connecting the $T_c$ points presents superconductivity observed in both the RF %%@
inductance and the $1/T_1$, and the dashed line presents superconductivity only seen in the RF inductance.}
\end{figure}

In this work, we present our $^{77}$Se NMR studies on high-quality FeSe single crystals with pressures up to 2.4 GPa and temperatures down to 50 mK. Our main results are summarized in %%@
the phase diagram of Fig.~\ref{Fig1}. From the pressure dependence of the NMR spectral splitting under an in-plane magnetic field, we observe a decrease of the structural transition %%@
temperature $T_s$ with pressure. However, the stripe-type spin fluctuations, characterized by the anisotropic 1/$^{77}T_1T$, are enhanced below a weakly pressure dependent temperature %%@
$T^*$ $\sim$ 100 K. We find that the magnetic ordering emerges about $\sim$ 0.2 GPa higher than earlier reports~\cite{Terashima_JPSJ_2015,SunJP_NC}.
The magnetic transition at $T_N$ above 2 GPa is first-order, and $T_N$ increases with pressure. The S-AFM phase necessarily breaks the $C_4$ symmetry as the magnetic ordering in iron %%@
pnictides. The discovery of the strong stripe-type spin fluctuations at high temperatures, the first-order magnetic transition, and the low-temperature S-AFM ground state under %%@
pressure clearly reveals a universal magnetic origin of the nematicity in both FeSe and iron pnictides. They also shed light on the important role of magnetism on superconductivity in %%@
iron-based superconductors.

Details of FeSe single crystal synthesis and characterization are presented in supplemental S1 and S2. The superconducting transition temperature $T_c$ is determined {\it in situ} by %%@
the RF inductance of the NMR coil (supplemental S3). The NMR measurements were performed under 10.3 T with two field configurations, $H \parallel a\&b$ (tetragonal [1 1 0] direction) %%@
and $H \parallel c$. Daphne oil was used as the pressure medium, and the pressure was determined by Cu$_2$O NQR resonant frequency at 5 K~\cite{Cu2O_NQR_HP}. For applying pressure %%@
above 2 GPa, the cell was heated to 80 $^{\circ}$C in order to improve pressure hydrostaticity~\cite{Yokogawa_JPSJ_2007}. Standard spin-echo and CPMG techniques were used for data %%@
accumulation to optimize the signal-to-noise ratio. The full-width-of-half-maximum (FWHM) of $^{77}$Se spectra is $\sim$ 3 kHz at 300 K at all pressures, indicating high sample %%@
quality and pressure homogeneity. The spin-lattice relaxation rates were measured with the inversion-recovery method, where perfect single-exponential functions with time are found %%@
with no signal loss or wipeout effect above $T_N$.

\begin{figure}[t]
\includegraphics[width=8.5cm, height=5.3cm]{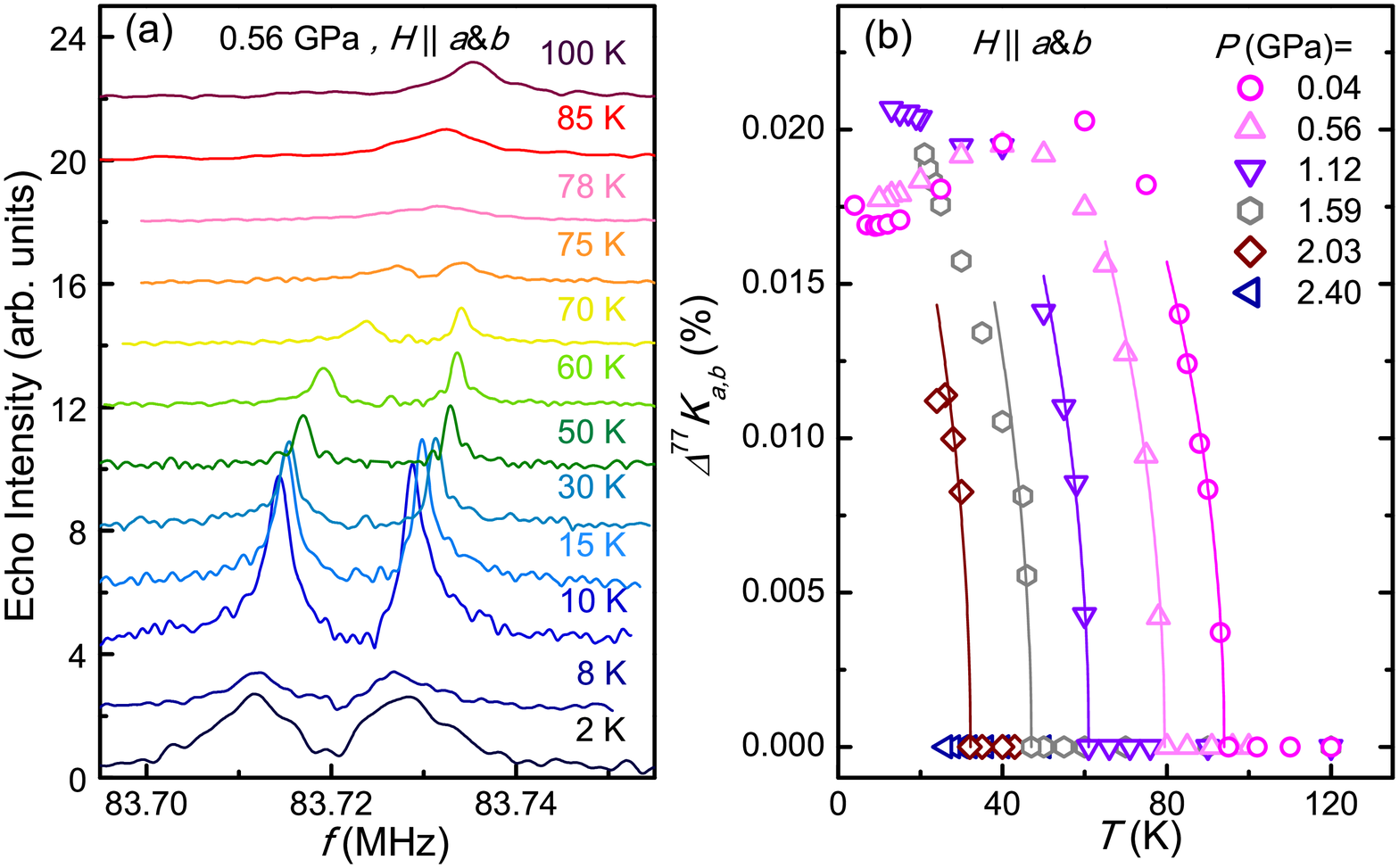}
\caption{\label{Fig2}(color online). (a) The characteristic $^{77}$Se spectra showing the structure transition by the line splitting with $H \parallel a\&b$ at $P$ = 0.56 GPa. (b) The %%@
difference of Knight shifts, $\Delta^{77}K_{a,b}=|^{77}K_a -^{77}K_b|$, measured as functions of temperature at various pressures. The solid lines are fits to mean-field functions %%@
$\Delta^{77}K_{a,b}\sim 1/(T_s-T)^{1/2}$.}
\end{figure}

\begin{figure}[t]
\includegraphics[width=8cm, height=9cm]{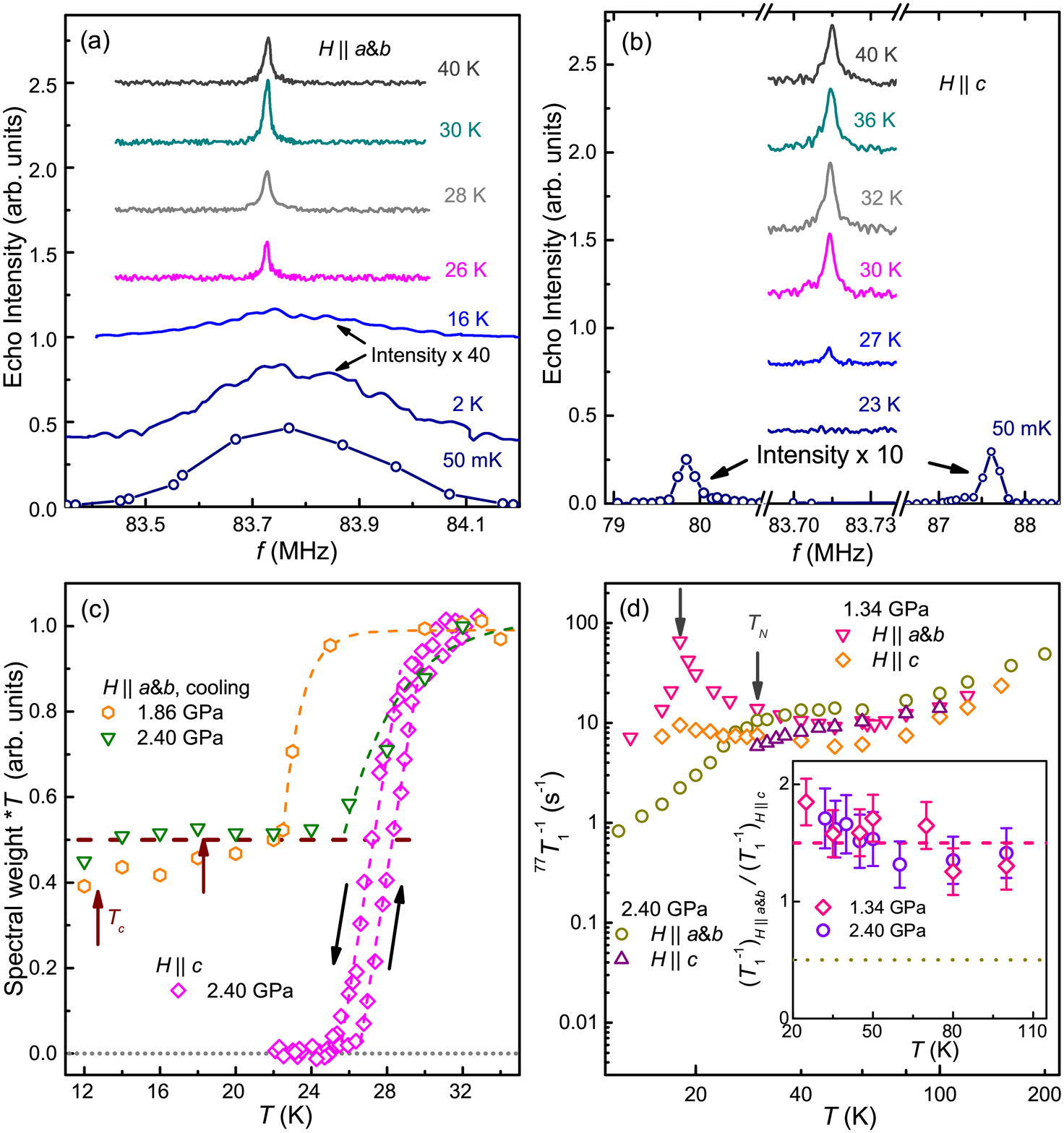}
\caption{\label{Fig3}(color online). (a) and (b): The NMR spectra with $H \parallel a\&b$ and $H \parallel c$ at $P$ = 2.4 GPa. (c) The normalized total spectral weight as a function %%@
of temperature at high pressures. $T_c$ are determined by the RF inductance under the same field. (d) The 1/$^{77}T_1$ for $H\parallel a\&b$ (averaged for two frequency peaks below %%@
$T_s$) and $H \parallel c$, at $P$ = 1.34 and 2.4 GPa. Inset: the anisotropy factor $R = (1/^{77}T_1)_{H\parallel a\&b}/(1/^{77}T_1)_{H \parallel c}$ as a function of temperature. The %%@
dashed and dotted horizontal lines are the theoretical values for the stripe-type ($R$ = 1.5) and for the checkerboard-type ($R$ = 0.5) spin fluctuations.}
\end{figure}

We first determine the evolution of the structural transition under pressure by tracing the NMR line splitting with $H \parallel a\&b$~\cite{Nat_Mater_NMR,PRL_C66_NMR}.
In the orthorhombic phase, the in-plane Knight shift is anisotropic. This leads to two resonance peaks
respectively corresponding to $H \parallel a$ and $H \parallel b$ due to structural twinning.
In Fig.~\ref{Fig2}(a), NMR spectra are shown at $P$ = 0.56 GPa for several selected temperatures, with evidence of structural transition from the NMR line splitting below $T_s$. The %%@
difference of Knight shifts, $\Delta^{77}K_{a,b} =|^{77}K_a -^{77}K_b|$, where $^{77}K_a$ and $^{77}K_b$ are the Knight shifts for $H \parallel a$ and $H \parallel b$ respectively, %%@
follows a mean-field like temperature dependence, consistent with a second-order phase transition~\cite{Nat_Mater_NMR}. Similar behavior of $\Delta^{77}K_{a,b}$ is observed with %%@
pressures up to $\sim$ 2 GPa (Fig.~\ref{Fig2}(b)). As presented in the phase diagram (Fig.~\ref{Fig1}), $T_s$ shows a gradual suppression with pressure below 2 GPa. At $P$ = 2.4 GPa, %%@
line splitting is absent above $T_N$ and cannot be resolved below $T_N$ (Fig.~\ref{Fig3}(a)).

\begin{figure*}[t]
\includegraphics[bb=0 0 1440 396, width=18cm, height=5cm]{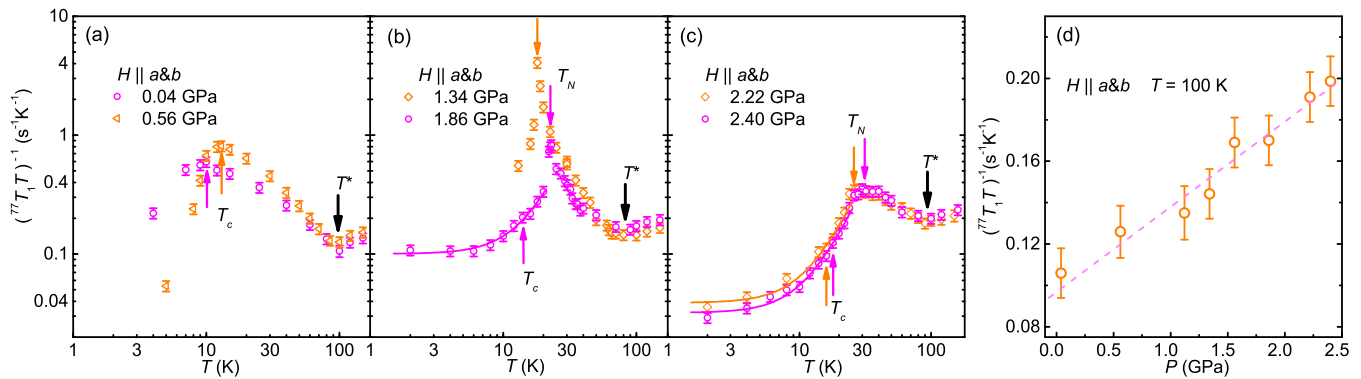}
\caption{\label{Fig4}(color online). (a), (b), (c): The spin-lattice relaxation rate divided by temperature ($1/^{77}T_1T$) with $H \parallel a\&b$ at typical pressures. Only the data %%@
measured on the high-frequency peaks are presented below $T_s$. The $T_c$ (determined by RF inductance under the same field), the $T_N$, and the $T^*$ are marked by the arrows. (d) %%@
The plot of $1/^{77}T_1T$ at 100 K as a function of pressure. }
\end{figure*}

Next we study the magnetic ordering in the pressurized phase. Below $T_s$, the spin-lattice relaxation rates, $1/^{77}T_1$, are different for two frequency peaks with $H \parallel %%@
a\&b$, and data are presented in Fig.~\ref{Fig4}(a)-(c) for the high-frequency peak and Fig. S5 for both peaks for comparison. The $1/^{77}T_1$ is larger for the high-frequency peak %%@
(Fig. S5), and the data for two frequencies at low pressures are consistent with the earlier report~\cite{Nat_Mater_NMR}. For $P$ $>$ 1.34 GPa, a magnetic phase transition is clearly %%@
seen from the peaked feature in $1/^{77}T_1T$ (Fig.~\ref{Fig4}(b)-(c) and S5) and the broadening of the NMR spectra upon cooling. Typical spectral data at 2.4 GPa are shown in %%@
Fig.~\ref{Fig3}(a)-(c). Below 30 K, the spectrum with $H \parallel a\&b$ shifts slightly to the high-frequency side, and its FWHM increases from $\sim$ 10 kHz at $T$ = 30 K to $\sim$ %%@
300 kHz at $T$ = 26 K (Fig.~\ref{Fig3}(a)), signaling the magnetic transition. Upon cooling, the spin-spin relaxation time $^{77}T_2$ increases from $\sim$ 5 ms at 120 K to $\sim$ 20 %%@
ms below $T_N$, which ensures correct analysis on the spectral weight. In Fig.~\ref{Fig3}(c), the normalized total spectral weight is drawn as a function of temperature at this %%@
pressure. The weight drops steeply by 50$\%$ from $T$ = 30 K to 26 K across the magnetic transition, develops a plateau-like feature between 26 K and 18 K, and then decreases slowly %%@
upon further cooling due to RF screening below $T_C$.

The $1/^{77}T_1$ shows a sudden drop at 28 K defined as $T_N$, coinciding with the middle point of the magnetic transition from the spectral loss. With $H \parallel c$, the spectra %%@
keep narrow (Fig.~\ref{Fig3}(b)), but the total spectral weight decreases when cooled below 30 K and reaches zero at 26 K (Fig.~\ref{Fig3}(c)). Two split broad NMR lines ($\sim$ 300 %%@
kHz) are seen at 50 mK (Fig.~\ref{Fig3}(b)), corresponding to $H_{in}^c \approx \pm $ 0.48 T.

The split spectra with $H \parallel c$ and the large spectral weight with $H \parallel a\&b$ resemble the $^{75}$As NMR spectra of stripe ordering in the iron %%@
pnictides~\cite{Commensur_BaFeAs_JPSJ}. By applying the same analysis on $^{77}$Se~\cite{Commensur_BaFeAs_JPSJ,T1_ratio_PRB}(supplemental S4), we show that the change of the spectral %%@
weight below $T_N$ is caused by the formation of a stripe-type AFM order. The ordered moment on Fe sites is projected as ($m_{Fe}^a$, $m_{Fe}^b$, $m_{Fe}^c$) to the principal axes of %%@
the orthorhombic structure, and the hyperfine field on the $^{77}$Se is calculated to be ($H_{in}^a$, $H_{in}^b$, $H_{in}^c$) = $A_{hf}^{ac}$($m_{Fe}^c$, 0, $m_{Fe}^a$), where %%@
$A_{hf}^{ac}$ is an off-diagonal hyperfine coupling constant. With a finite $\pm m_{Fe}^a$ ($\pm H_{in}^c$), the spectrum splits for $H \parallel c$ but does not split for $H %%@
\parallel b$. This produces exactly what we have observed in the ordered phase: with $H \parallel c$ only symmetric NMR lines are observed $\sim$ 3.9 MHz away from the center, whereas %%@
with $H \parallel b$ spectral weight remains near the paramagnetic frequency.

The disappearance of the paramagnetic peak with $H\parallel c$ indicates that the sample is fully magnetically ordered. The 50$\%$ signal loss with $H \parallel a\&b$ could be caused %%@
by a broad spectrum from distributed magnetic moments, or a very short $T_2$ from phase inhomogeneity.
Moreover, we find that the signal loss remains 50$\%$ for different pressures and samples, and another plausible explanation is
that one of the two domains is out of the NMR window in the stripe phase. In particular, if $m_{Fe}^c$ ($H_{in}^a$) is finite, the signal is lost for $H \parallel a$ but not for $H %%@
\parallel b$. However, further experimental evidences and theoretical understanding are requested to fully settle this scenario.

The above analyses already allow us to rule out other proposed local patterns, such as the checkerboard ($\pi$, $\pi$) spin orders where a zero $c$-axis internal field is expected on %%@
the Se sites (supplemental S4). In fact, it has been suggested theoretically that the lack of magnetic ordering at the ambient pressure is caused by competing interactions, such as %%@
strong magnetic frustration from nearest and next nearest neighbor exchange couplings $J_1$ and  $J_2$~\cite{PRL_RongYu,Nat_Phy_WangFa,Nat_Phy_Frustrat,LiuK_PRB_2016}. Our observation %%@
under pressure puts strong constrains on these competing theories: naively, our finding of the S-AFM phase suggests a reduced ratio of $J_1/J_2$ with pressure. This is also consistent %%@
with an $ab$ $initio$ DFT calculation, where the S-AFM state has the lowest energy over other magnetic states under pressure~\cite{Nat_Phy_Frustrat}.

This high-pressure magnetic order of FeSe naturally resembles that of the iron pnictides~\cite{P_Dai_Nat}, and therefore may fit to a unified magnetic phase diagram of the iron-based %%@
superconductors generated by the competing exchange interactions. The stripe order with a finite $m_{Fe}^a$ clearly indicates a magnetic $C_4$ symmetry breaking because of two choices %%@
to select $a$-axis of the crystals. Furthermore, a strong spin-lattice coupling has to be considered in the magnetic phase transition as shown below. We examined the magnetic %%@
transition across $T_N$ from the temperature scan of the $1/^{77}T_1T$, as shown in Fig.~\ref{Fig4}(a)-(c) for the high-frequency peak. In fact, the following discussions are valid %%@
for the low-frequency peak as well (see Fig. S5). For pressures from 1.34 GPa to 1.86 GPa, the $1/^{77}T_1T$ has a clear divergence at $T_N$, suggesting a second-order magnetic %%@
transition. Surprisingly, the divergence is completely absent at $T_N$ under higher pressures ($\geq$ 2 GPa), directly evidencing a strong first-order transition with no critical %%@
fluctuations. The first-order nature of the transition is further supported by the hysteresis near $T_N$, with a 1.5 K shift in the temperature dependence of the integrated spectral %%@
weight with $H \parallel c$ between cooling and warming at 0.1 K/minute (Fig.~\ref{Fig3}(c)). This strongly implies that the magnetic transition is coupled to a %%@
tetragonal-to-orthorhombic structural transition due to the strong spin-lattice coupling~\cite{YeQ_PRB_2013,Lischner_PRB_2015}. The simultaneous magnetic and structural
transition are also supported by the high-pressure XRD data~\cite{HP_lattice_para,HP_Xray_NC}. It also suggests that the structural transition is an Ising-nematic transition with a %%@
magnetic origin~\cite{Fang_PRB_2008,Dai_PNAS_2009}.

We are now in a position to discuss the spin fluctuations above $T_N$ from the $1/^{77}T_1$ data. Fig.~\ref{Fig3}(d) shows the $(1/^{77}T_1)_{H \parallel a\&b}$ and $(1/^{77}T_1)_{H %%@
\parallel c}$ below 200 K at $P$ = 1.34 GPa and 2.4 GPa. In the inset, the anisotropy factor $R = (1/^{77}T_1)_{H \parallel a\&b}/(1/^{77}T_1)_{H \parallel c}$ at these two pressures %%@
are presented, where $R \approx 1.5$ holds from 100 K down to 40 K. This value of $R$ is an indication of stripe-type, or ($\pi$, 0), spin fluctuations at temperatures far above $T_N$ %%@
(supplemental S4). $R \approx 1.5$ persists to the highest pressure we measured, and is consistent with the ground state magnetism we presented. The possibility of checkerboard-type %%@
spin fluctuations~\cite{ZhaoJun_arxiv_1511} with $R \approx 0.5$ (see Fig.~\ref{Fig3}(d) inset) is ruled out at high pressures.

Remarkably, as shown in Fig.~\ref{Fig4}(a)-(c), the high-temperature $1/^{77}T_1T$ first decreases upon cooling below 200 K, and then shows an upturn behavior with further decrease of %%@
temperature. For each pressure, we define a characteristic temperature $T^*$ for the onset of the upturn [as shown in Fig.~\ref{Fig4}(a)-(c)], which indicates enhanced low-energy spin %%@
fluctuations. The $T^*$ barely varies with pressure within our resolution, as shown in the phase diagram (Fig.~\ref{Fig1}). The error bars of $T^*$ are taken as distances between two %%@
temperatures where $1/^{77}T_1T$ is enhanced by 10{\%} when cooling/warming away from $T^*$. Furthermore, the $1/^{77}T_1T$ at 100 K, as presented in Fig.~\ref{Fig4}(d), also %%@
increases with pressure, consistent with earlier NMR results on polycrystals~\cite{Imai_PRL_2009}. Therefore, our high-pressure data reconcile FeSe and other iron-based %%@
superconductors, where the stripe order and the ($\pi$, 0) fluctuations tend to be universal, although distinct properties are observed in FeSe at the ambient pressure.

These high-pressure data shed important new light on the driving force of the nematicity. First, in the full range of pressure we have measured, enhanced low-energy stripe-type spin %%@
fluctuations exist up to $T^*$. $T^*$ only accidentally coincides with $T_s$ (orbital ordering) at $P=0$, but surpasses $T_s$ largely at high pressures. The emergence of %%@
high-temperature stripe-type spin fluctuations is hardly understood within an orbital-driven-nematicity scenario~\cite{Nat_Mater_NMR}, but could be explained as enhanced spin %%@
fluctuations above the Ising-nematic transition in a magnetic-driven-nematicity scenario~\cite{Rong_PRB_2012}. Second, the concomitant structural transition above 2 GPa, manifest by %%@
the first-order magnetic phase transition, evidences a strong coupling between nematicity and magnetism under pressure. This scenario implies the same underlying physics governing the %%@
coupled magnetic and nematic transitions in some iron pnictides~\cite{Kasahara_Nature}, and hence suggests a universal picture of a magnetic-driven nematicity in iron-based %%@
superconductors~\cite{Rev_nematic}. Furthermore, it has been proposed that frustrated magnetic exchange interactions at low pressures may favor other magnetic orders, such as the %%@
antiferroquadrupolar order~\cite{PRL_RongYu} and/or the staggered dimmer/trimmer order~\cite{Nat_Phy_Frustrat, LiuK_PRB_2016}. These exotic magnetic states support the same nematic %%@
order, but compete with the stripe-order magnetism. With increasing pressure, these competing orders may be suppressed while the stripe correlations grow strongly, which lead to a %%@
non-monotonic change of $T_s$~\cite{PRL_RongYu,Nat_Phy_Frustrat} as we observed.

Finally, we address the implications of our data to superconductivity. Besides the RF inductance measurements (supplemental S3), the onset of superconductivity below 1 GPa is also %%@
shown by a kinked feature in the $1/^{77}T_1T$ upon cooling (Fig.~\ref{Fig4}(a)), which signals bulk superconductivity. Above 1.5 GPa, while both $T_c$ and $T_N$ increase with %%@
pressures, the $1/^{77}T_1T$ drops smoothly and fits to $1/^{77}T_1T=a+bT^\alpha$ with $\alpha\approx 2.5$ (presented by the solid lines in Fig.~\ref{Fig4}(b-c)),
which is a typical form by taking into account the contributions from both itinerant electrons and spin waves far below $T_N$.
The absence of a kinked feature across $T_c$ in $1/^{77}T_1T$ excludes the microscopic coexistence of superconductivity in our observed stripe phase. However, further investigation is %%@
needed to address the exact locations and the properties of the superconducting phase. Since superconductivity does not show up in the observed magnetic regions, we speculate that it %%@
exists in small inhomogeneous or short $T_2$ regions as we described earlier. Nevertheless, the proximity of the superconducting phase to the stripe order and existence of strong %%@
($\pi$, 0) spin fluctuations in the paramagnetic phase, draw a close relation between superconductivity and the stripe-order magnetism, as seen in iron pnictides. Future studies under %%@
higher pressures, when magnetic ordering is suppressed~\cite{SunJP_NC}, may shed further light on the pairing mechanism~\cite{Essenberger_arxiv,T.W_PRX_2015}.

In summary, we report direct spectroscopic evidence for the strong suppression of the orbital ordering/structure transition under pressure, and for a stripe-order magnetism above 2 %%@
GPa in FeSe. The magnetic transition is identified as a first-order type with a $C_4$ symmetry breaking. These pressure effects put constraints on theories of magnetism in FeSe. %%@
Although the $T_S$ is not directly detected by the current NMR data above 2 GPa, electronic nematicity and nematic fluctuations are shown to be closely coupled to the stripe-order %%@
magnetism, resulting in a first-order magnetic and structural transition under high pressures and persistent strong ($\pi$, 0) spin fluctuations over a wide range of temperature and %%@
pressure. These results suggest a strong coupling among lattice, magnetism, and nematicity, and fully support a magnetic-driven-nematicity scenario. Our high-pressure data also %%@
reconcile FeSe with other iron-based superconductors by the stripe-order magnetism, and helps to understand the superconductivity on a universal basis.

We acknowledge encouraging discussions with Zhong-Yi Lu, Dong-Hai Lee, Tao Xiang, Kai Liu, Fa Wang, and Qimiao Si. Work at Renmin University of China is supported by the National %%@
Science Foundation of China (NSFC) (Grant Nos. 11222433, 11374361, 11374364 and 11574394), the Ministry of Science and Technology of China (Grant Nos. 2016YFA0300504), and the %%@
Fundamental Research Funds for the Central Universities and the Research Funds of Renmin University of China (Grant Nos. 14XNLF08, 15XNLQ07 and 15XNLF06).


\begin{thebibliography}{38}

\bibitem{NatP_6_645}
J. Paglione, and R. L. Greene,
%High-temperature superconductivity in iron-based materials.
\newblock Nature Phys. \textbf{6}, 645 (2010).

\bibitem{Stewart_RevMP}
G. R. Stewart,
%Superconductivity in Iron Compounds.
\newblock Rev. Mod. Phys. \textbf{83}, 1589 (2011).


\bibitem{P_Dai_RevMP}
P. Dai,
\newblock Rev. Mod. Phys. \textbf{87}, 855 (2015).



\bibitem{P_Dai_Nat}
C. de la Cruz, Q. Huang, J. W. Lynn, J. Li, W. Ratcliff II, J. L. Zarestky, H. A. Mook, G. F. Chen, J. L. Luo, N. L. Wang, and P. Dai,
%Magnetic order close to superconductivity in the iron-based layered LaO1-xFxFeAs %systems.
\newblock Nature \textbf{453}, 899 (2008).

\bibitem{Science_329_824}
J.-H. Chu, J. G. Analytis, K. De Greve, P. L. McMahon, Z. Islam, Y. Yamamoto and I. R. Fisher,
%\newblock In-plane resistivity anisotropy in an underdoped iron arsenide
%  superconductor.
\newblock Science \textbf{329},824 (2010).


\bibitem{Yi11}
M. Yi, D. Lu, J.-H. Chu, J. G. Analytis, A. P. Sorini, A. F. Kemper, B. Moritz, S.-K. Mo, R. G. Moore, M. Hashimoto, W.-S. Lee, Z Hussain, T. P. Devereaux, I. R. Fisher, and Z.-X. %%@
Shen,
%Symmetry-breaking orbital anisotropy observed for detwinned Ba(Fe1-xCox)2As2 above the spin density wave %transition.
\newblock Proc. Natl. Acad. Sci. \textbf{108}, 6878 (2011).


\bibitem{Lv}
W. Lv, F. Kr\"{u}ger, and P. Phillips, Phys. Rev. B \textbf{82}, 045125 (2010).




\bibitem{Rev_nematic}
R. M. Fernandes, A. V. Chubukov, and J. Schmalian,
%What drives nematic order in iron-based superconductors?
\newblock Nature Phys. \textbf{10}, 97 (2014).

\bibitem{WuMK_PNAS_2008}
F.-C. Hsu, J.-Y. Luo, K.-W. Yeh, T.-K. Chen, T.-W. Huang, P. M. Wu, Y.-C. Lee, Y.-L. Huang, Y.-Y. Chu, D.-C. Yan, and M.-K. Wu,
%Superconductivity in the PbO-type structure ¦Á-FeSe.
\newblock Proc. Natl. Acad. Sci. \textbf{105}, 14262 (2008).

\bibitem{Cava_PRL_2009}
T. M. McQueen, A. J. Williams, P. W. Stephens, J. Tao, Y. Zhu, V. Ksenofontov, F. Casper, C. Felser, and R. J. Cava,
%Tetragonal-to-Orthorhombic Structural Phase Transition at 90 K in the %Superconductor Fe1.01Se.
\newblock Phys. Rev. Lett. \textbf{103}, 057002 (2009).

\bibitem{Shimojima_PRB}
T. Shimojima, Y. Suzuki, T. Sonobe, A. Nakamura, M. Sakano, J. Omachi, K. Yoshioka, M. Kuwata-Gonokami, K. Ono, H. Kumigashira, A. E. B\"ohmer, F. Hardy, T. Wolf, C. Meingast, H. v. %%@
L\"ohneysen, H. Ikeda, and K. Ishizaka,
%Lifting of xz/yz orbital degeneracy at the structural transition in detwinned FeSe.
\newblock Phys. Rev. B \textbf{90}, 121111(R) (2014).


\bibitem{Imai_PRL_2009}
T. Imai, K. Ahilan, F. L. Ning, T. M. McQueen, and R. J. Cava,
%Why Does Undoped FeSe Become a High-Tc Superconductor under Pressure?
\newblock Phys. Rev. Lett. \textbf{102}, 177005 (2009).


\bibitem{Bendele_uSR_PRL_2010}
M. Bendele, A. Amato, K. Conder, M. Elender, H. Keller, H.-H. Klauss, H. Luetkens, E. Pomjakushina, A. Raselli, and R. Khasanov,
%Pressure Induced Static Magnetic Order in Superconducting FeSe1-x.
\newblock Phys. Rev. Lett. \textbf{104}, 087003 (2010).


\bibitem{Terashima_JPSJ_2015}
T. Terashima, N. Kikugawa, S. Kasahara, T. Watashige, T. Shibauchi, Y. Matsuda, T. Wolf, Anna E. B\"ohmer, Fr\'ed\'eric Hardy, C. Meingast, H. v. L\"ohneysen, and S. Uji,
%Pressure-Induced Antiferromagnetic Transition and Phase Diagram in FeSe.
\newblock J. Phys. Soc. Jpn. \textbf{84}, 063701 (2015).

\bibitem{SunJP_NC}
J. P. Sun, K. Matsuura, G. Z. Ye, Y. Mizukami, M. Shimozawa, K. Matsubayashi, M. Yamashita, T. Watashige, S. Kasahara, Y. Matsuda, J.-Q. Yan, B. C. Sales, Y. Uwatoko, J.-G. Cheng, and %%@
T. Shibauchi,
%Dome-shaped magnetic order competing with high-temperature superconductivity at %high pressures in FeSe.
\newblock Nat. Commun. \textbf{7}, 12146 (2016).

\bibitem{PRL_RongYu}
R. Yu and Q. Si,
%Antiferroquadrupolar and Ising-Nematic Orders of a Frustrated Bilinear-Biquadratic %Heisenberg Model and Implications for the Magnetism of FeSe.
\newblock Phys. Rev. Lett. \textbf{115}, 116401 (2015).


\bibitem{Nat_Phy_WangFa}
F. Wang, S. A. Kivelson, and D.-H. Lee,
%Nematicity and quantum paramagnetism in FeSe.
\newblock Nature Phys. \textbf{11}, 959 (2015).

\bibitem{Nat_Phy_Frustrat}
J. K. Glasbrenner, I. I. Mazin,	Harald O. Jeschke, P. J. Hirschfeld, R. M. Fernandes, and Roser Valent\'i,
%E.ect of magnetic frustration on nematicity and superconductivity in iron %chalcogenides.
\newblock Nature Phys. \textbf{11}, 953 (2015).

\bibitem{LiuK_PRB_2016}
K. Liu, Z.-Y. Lu, and T. Xiang,
%Nematic antiferromagnetic states in bulk FeSe.
\newblock Phys. Rev. B \textbf{93}, 205154 (2016).


\bibitem{Rahn_PRB_2015}
M. C. Rahn, R. A. Ewings, S. J. Sedlmaier, S. J. Clarke, and A. T. Boothroyd,
%Strong (¦Ð,0) spin fluctuations in ¦Â-FeSe observed by neutron spectroscopy.
\newblock Phys. Rev B 91, 180501(R) (2015).

\bibitem{ZhaoJun_arxiv_1511}
Q. Wang, Y. Shen, B. Pan, X. Zhang, K. Ikeuchi, K. Iida, A. D. Christianson, H. C. Walker, D. T. Adroja, M. Abdel-Hafiez, Xiaojia Chen, D. A. Chareev, A. N. Vasiliev, and J. Zhao,
%Magnetic ground state of FeSe.
\newblock Nat. Commun. \textbf{7}, 12182 (2016).

\bibitem{Garbarino_34K_EPL}
G. Garbarino, A. Sow, P. Lejay, A. Sulpice, P. Toulemonde, M. Mezouar, and M. N\'u\~nez-Regueiro,
%High-temperature superconductivity (Tc onset at 34 K) in the high-pressure %orthorhombic phase of FeSe.
\newblock Europhys. Lett. \textbf{86}, 27001 (2009).

\bibitem{Medvedev_NatMat_HP}
S. Medvedev, T. M. McQueen, I. A. Troyan, T. Palasyuk, M. I. Eremets, R. J. Cava, S. Naghavi, F. Casper, V. Ksenofontov, G. Wortmann, and C. Felser,
%Electronic and magnetic phase diagram of ¦Â-Fe1.01Se with superconductivity at 36.7 %K under pressure
\newblock Nature Mater. \textbf{8}, 630 (2009).


\bibitem{Cu2O_NQR_HP}
A. P. Reyes, E. T. Ahrens, R. H. Heffner, P. C. Hammel, and J. D. Thompson,
\newblock Rev. Sci. Instrum. \textbf{63}, 3120 (1992).

\bibitem{Yokogawa_JPSJ_2007}
K. Yokogawa, K. Murata, H. Yoshino, and S. Aoyama,
\newblock Jpn. J. Appl. Phys. \textbf{46}, 2626 (2007).

\bibitem{Nat_Mater_NMR}
S-H. Baek, D. V. Efremov, J. M. Ok,	J. S. Kim, Jeroen van den Brink, and B. B\"uchner,
%Orbital-driven nematicity in FeSe.
\newblock Nature Mater. \textbf{14}, 210 (2015).


\bibitem{PRL_C66_NMR}
A. E. B\"ohmer, T. Arai, F. Hardy, T. Hattori, T. Iye, T. Wolf, H.v. L\"ohneysen, K. Ishida, and C. Meingast,
%Origin of the Tetragonal-to-Orthorhombic Phase Transition in FeSe: A Combined %Thermodynamic and NMR Study of Nematicity.
\newblock Phys. Rev. Lett. \textbf{114}, 027001 (2015).

\bibitem{Commensur_BaFeAs_JPSJ}
K. Kitagawa, N. Katayama, K. Ohgushi, M. Yoshida, and M. Takigawa,
\newblock J. Phys. Soc. Jpn. \textbf{77}, 114709 (2008).

\bibitem{T1_ratio_PRB}
S. Kitagawa, Y. Nakai, T. Iye, K. Ishida, Y. Kamihara, M. Hirano, and H. Hosono,
\newblock Phys. Rev. B \textbf{81}, 212502 (2010).

\bibitem{YeQ_PRB_2013}
Q. Q. Ye, K. Liu, and Z. Y. Lu,
%Influence of spin-phonon coupling on antiferromagnetic spin fluctuations in FeSe %under pressure: First-principles calculations with van der Waals corrections.
\newblock Phys. Rev. B \textbf{88}, 205130 (2013).

\bibitem{Lischner_PRB_2015}
J. Lischner, T. Bazhirov, A. H. MacDonald, M. L. Cohen, and S. G. Louie,
%First-principles theory of electron-spin fluctuation coupling and superconducting %instabilities in iron selenide.
\newblock Phys. Rev. B \textbf{91}, 020502(R) (2015).

\bibitem{HP_lattice_para}
S. Margadonna, Y. Takabayashi, Y. Ohishi, Y. Mizuguchi, Y. Takano, T. Kagayama, T. Nakagawa, M. Takata, and K. Prassides,
%Pressure evolution of the low-temperature crystal structure and bonding of the %superconductor FeSe (Tc=37 K)
\newblock Phys. Rev. B \textbf{80}, 064506 (2009).

\bibitem{HP_Xray_NC}
K. Kothapalli, A. E. B\"ohmer, W. T. Jayasekara, B. G. Ueland, P. Das, A. Sapkota, V. Taufour, Y. Xiao, E. Alp, S. L. Bud'ko, P. C. Canfield, A. Kreyssig, and A. I. Goldman,
%Strong cooperative coupling of pressure-induced magnetic order and nematicity in FeSe
\newblock Nat. Commu. \textbf{1}, 12728 (2016).

\bibitem{Fang_PRB_2008}
C. Fang, H. Yao, W.-F. Tsai, J. P. Hu, and S. A. Kivelson,
%Theory of electron nematic order in LaFeAsO.
\newblock Phys. Rev. B \textbf{77}, 224509 (2008).

\bibitem{Dai_PNAS_2009}
J. Dai, Q. Si, J.-X. Zhu, and E. Abrahams,
%
\newblock Proc. Natl. Acad. Sci. \textbf{106}, 4418 (2009).


\bibitem{Rong_PRB_2012}
R. Yu, Z. Wang, P. Goswami, A. H. Nevidomskyy, Q. Si, and E. Abrahams,
%Spin dynamics of a J1-J2-K model for the paramagnetic phase of iron pnictides.
\newblock Phys. Rev. B \textbf{86}, 085148 (2012).

\bibitem{Kasahara_Nature}
S. Kasahara, H. J. Shi, K. Hashimoto, S. Tonegawa, Y. Mizukami, T. Shibauchi, K. Sugimoto, T. Fukuda, T. Terashima, Andriy H. Nevidomskyy, and Y. Matsuda,
%Electronic nematicity above the structural and superconducting transition in
%BaFe2(As1-xPx)2
\newblock Nature \textbf{486}, 382 (2012).

\bibitem{Essenberger_arxiv}
F. Essenberger, A. Sanna, P. Buczek, A. Ernst, L. Sandratskii, and E.K.U. Gross,
%Ab-initio theory of Iron based superconductors.
\newblock arXiv:1411.2121.

\bibitem{T.W_PRX_2015}
T. Watashige, Y. Tsutsumi, T. Hanaguri, Y. Kohsaka, S. Kasahara, A. Furusaki, M. Sigrist, C. Meingast, T. Wolf, H. v. L\"ohneysen, T. Shibauchi, and Y. Matsuda,
%Evidence for Time-Reversal Symmetry Breaking of the Superconducting State near %Twin-Boundary Interfaces in FeSe Revealed by Scanning Tunneling Spectroscopy.
\newblock Phys. Rev. X \textbf{5}, 031022 (2015).


\end{thebibliography}
\end{document}